\documentclass{elsart}

\usepackage{graphicx}
\usepackage{epsfig}

\usepackage{amssymb}

\begin{document}

\begin{frontmatter}

\title{Breaking records in the evolutionary race} 

\author{Joachim Krug and Kavita Jain}

\address{Institut f\"ur Theoretische Physik, Universit\"at zu 
K\"oln \\ Z\"ulpicher Strasse 77, 50937 K\"oln, Germany}

\begin{abstract}

We explore some aspects of the relationship between biological evolution
processes and the mathematical theory of records. 
For Eigen's quasispecies model with an uncorrelated fitness landscape, we show that
the evolutionary trajectories traced out by a population initially localized at a randomly
chosen point in sequence space can be described in close analogy
to record dynamics, with two complications.
First, the increasing number of genotypes that become available 
with increasing distance from the starting point implies that fitness records are
more frequent than for the standard case of independent, identically
distributed random variables. Second, fitness records can be bypassed,
which strongly reduces the number of genotypes that take part in an evolutionary 
trajectory. For exponential and Gaussian fitness distributions, this number
scales with sequence length $N$ as $\sqrt{N}$, 
and it is of order unity for distributions with a power law tail.
This is in strong contrast to the number of records, which is of order $N$
for any fitness distribution.

\end{abstract}

\begin{keyword} Biological evolution \sep punctuated equilibrium \sep
record dynamics \sep
extremal statistics \sep quasispecies model 

\end{keyword}
\end{frontmatter}

\section{Records and evolution}
\label{Records}

In the Darwinian view of nature \cite{Darwin}, biological evolution is a fierce
competition among different organisms in which the winners
are rewarded by copious offspring while the losers perish. 
It should therefore be no surprise to see metaphors from the world of 
athletics turn up in the description of evolutionary dynamics. Indeed,
every evolutionary innovation that is fixed in a population has
to be a \textit{record}, in the sense that it solves 
some problem encountered by the organism in a way that is
superior to all existing solutions.
A possible mathematical relationship between evolution models and the 
theory of records was suggested by Kauffman and Levin
in the context of long-jump adaption on correlated fitness landscapes 
\cite{Kauffman87}, and has more recently been elaborated by Sibani
and coworkers \cite{Sibani98}. 

The basic problem
of record statistics can be formulated as follows 
\cite{Glick78,Nevzorov87}: Given an ordered
sequence $\{ X_n \}_{n=1,2,3...}$ of real random variables (RV's), 
a \emph{record} occurs at $r$ iff 
\begin{equation}
\label{record}
X_r = \max_{n \leq r}\{ X_n \}.
\end{equation}
By convention, $X_1$ is always a record, and through application of 
(\ref{record}) a series of \emph{record times} $\{ r_k \}_{k=1,2,3...}$
and \emph{record values} $\{ X_{r_k} \}$ is generated from the underlying
sequence $\{ X_n \}$, with $r_k$ denoting the time of the
$k$'th record and $r_1 = 1$. 
Many properties of records are known for the case
when the $X_n$ are independent and identically distributed (i.i.d.). 
In particular, the statistics of the record times is completely independent
of the underlying probability distribution. 
This is largely a consequence of a simple symmetry argument \cite{Feller71}. Denote by
$P_n$ the probability that a record occurs at $n$. In the i.i.d. case,
each of the $n$ RV's $\{ X_1, X_2,...,X_n \}$ is equally likely to be
the largest, and hence $P_n = 1/n$. In particular, the expected number of records
$\langle R (n) \rangle$ up to time $n$ is equal to 
\begin{equation}
\label{log}
\langle R(n) \rangle = \sum_{i=1}^n  \frac{1}{i} \approx \ln (n) + 
0.57721566... + O(1/n).
\end{equation}
The full distribution of $R(n)$ becomes
Poissonian for large $n$, and the record sequence can be described
as a Poisson process in logarithmic time $\ln (n)$ \cite{Sibani98}.     
Furthermore, it can be shown that the ratios of subsequent
record times $r_k/r_{k+1}$ become uniformly distributed, independent
random variables for large $k$ \cite{Glick78}. This implies that the sequence
$\{ r_k \}$ of record times has some rather counterintuitive properties;
for example, given the time $r_k$ of the $k$'th record, the expected
time of the preceding record is $\langle r_{k-1} \rangle = r_k/2$, while
the expected time $\langle r_{k+1} \rangle $ of the next record is infinite\footnote{The latter 
property invalidates an argumentation based on the average waiting
time for the next record, which has lead Kauffman and Levin to conclude
(erroneously) that the number of records grows as $\log_2(n)$ rather than
as $\ln(n)$ \cite{Kauffman87}.}.       

The record sequence is distinctly non-stationary: 
With increasing time, it becomes exponentially harder to beat
the current record. For this reason record dynamics and the 
associated log-Poisson process has been invoked to describe the nonstationary aspects of macroevolutionary
dynamics \cite{Sibani98} (as evidenced e.g. by extinction and origination rates
of taxa in the fossil record \cite{Newman99}), as well as the relaxation of disordered
systems such as spin glasses \cite{Sibani03}. The pattern of static periods of
exponentially increasing duration 
interspersed by rare events of rapid change (new records) is a simple
realization of \textit{punctuated equilibrium}, an important paradigm of evolutionary
theory \cite{Gould93,Eldredge89}. 
        
Here we approach the relation between evolution and records from the 
point of view of population dynamics on the space of genetic sequences 
\cite{Krug02,Krug03}. We show how the properties of sequence space introduce
modifications to the standard record problem, which are of interest in their
own right, and only partly understood at present.  
Some basic notions are introduced in the next section,
and the remaining sections summarize the main results of a detailed
investigation, which will be published elsewhere \cite{Jain04}. 

\section{Sequence space and fitness landscape}

The proper arena in which to describe evolutionary dynamics is the space of 
genotypes, which are represented as sequences $\sigma = (\sigma_1, \sigma_2,...,
\sigma_N)$ of $N$ symbols taken from an alphabet of $\ell$ letters; for
DNA sequences $\ell = 4$, but in many theoretical studies binary sequences
($\ell = 2$) are considered for simplicity. The total number of possible
sequences is $S = \ell^N$. The nearest neighbors of a given
sequence $\sigma$ are those sequences $\sigma'$ that can be reached from 
$\sigma$ by a single point mutation, which alters one of the $N$ symbols.
More generally, the \textit{Hamming distance} $d(\sigma,\sigma')$ beween two
sequences $\sigma$ and $\sigma'$ is the number of symbols
in which the two differ. An important quantity in what follows is the number $\alpha_k$ of 
sequences at distance $k$ from a given sequence, which takes the form
\begin{equation}
\label{alphak}
\alpha_k = {N \choose k} \; (\ell-1)^{k}.
\end{equation}
This can be derived by noting that there are ${N \choose k}$ ways of choosing
$k$ mutation sites on the sequence, and at each site $\ell - 1$ different symbols
are available. The maximum distance between two sequences is $N$. For large $N$
(\ref{alphak}) takes the form of a Gaussian of width $\sqrt{N}$ centered around
the distance $k_{\mathrm{max}} = N (\ell-1)/\ell$ at which the majority of sequences 
reside. 

Next we have to associate a fitness
with each sequence $\sigma$. 
We define the fitness $W(\sigma)$, in the Wrightian sense \cite{Baake99}, to be
\textit{proportional to the expected number of offspring of an individual carrying the genotype 
$\sigma$} \cite{Peliti95,Peliti97}. This implies that  
$W(\sigma) \geq 0$, and only ratios of fitnesses matter. We can thus write
$W(\sigma) = e^{\beta F(\sigma)}$ to introduce an 
\textit{inverse selective temperature} $\beta$ \cite{Peliti97}
for later use. In the following both $W$ and $F$ will be referred to as ``fitness''.  

The mapping from genotype to fitness is largely unknown,
but it is expected to be very complicated. We therefore follow a common practice
and assume the $F(\sigma)$ to be quenched i.i.d. RV's drawn from some
distribution $p(F)$; in statistical physics
this is known as the \emph{random energy model} (REM) of spin glasses 
\cite{Peliti95,Franz93}, 
while in the context of evolutionary
biology it has been referred to as the \emph{house of cards model} \cite{Baake99}
or the \emph{uncorrelated fitness landscape} \cite{Kauffman87}.
Many properties of the REM fitness landscape, such as the number
of local fitness maxima 
and the length of uphill adaptive walks \cite{Kauffman87,Flyvbjerg92}, can be derived
using simple ideas from order statistics \cite{David70}. 
It is of particular interest
to find properties that are independent of the fitness distribution.
For example, the probability that a given sequence is a local maximum is
equal to the probability that it has
the largest fitness in the set of sequences comprising its $(\ell-1)N$
nearest neighbors and itself; by the symmetry argument 
of Sect.\ref{Records}, this is just $[(\ell-1)N+1]^{-1}$. 

Important characteristics of the REM landscape needed in the following discussion
are the expected maximum fitness value
$F_{\mathrm{max}}(S)$ that occurs among the $S$ independent sequences,
and the \emph{fitness gap} $\epsilon$, which is the difference
between the largest and the second largest fitness value \cite{Krug02}.
A simple estimate for the maximum fitness is obtained by setting  
the cumulative fitness distribution $p_c(F)$
equal to $1 - 1/S$ \cite{Sornette00}, 
\begin{equation}
\label{Fmax}
p_c(F_{\mathrm{max}}) =  \int_{-\infty}^{F_{\mathrm{max}}} dF \; p(F) = 1 - 1/S,
\end{equation}  
and the fitness gap is of the order of $\epsilon \sim [S p(F_{\mathrm{max}}(S))]^{-1}$
\cite{Krug02,Krug03}.

\section{Records in sequence space}
\label{RecordsSeq}

Kauffman and Levin \cite{Kauffman87} found record statistics to be applicable
in a situation where a population, assumed to be localized at a single
sequence at all times, explores sequence space by random mutations of arbitrarily
long range, and moves to a new location whenever the fitness of the mutant exceeds
that of the present position. 
To highlight the role of the geometry of sequence space, we consider here
a variant of their model where the range of mutations is restricted
but grows in the course of time. At time $t=0$ the 
entire population resides at a randomly chosen ``seed'' sequence $\sigma_0$. At 
the integer time $t > 0$, the population has access to all genotypes within
Hamming distance $k=t$ of $\sigma_0$, and it always resides in its entirety
at the sequence of maximum fitness within the accessible region. Thus
the current position of the population in sequence space \emph{jumps} whenever
a fitness record occurs among the $\alpha_k$ sequences which become
newly available at time $t=k$. 

The analysis of this model requires a slight generalization of the basic
symmetry argument of record statistics outlined above, which is adapted
to a situation where a variable number of new i.i.d. RV's 
is introduced at each time step\footnote{This generalization was originally
introduced to investigate whether the frequent breaking of records
in the Olympic games can be attributed to the fact that the athletes are
selected from exponentially growing populations \cite{Yang75}.
The conclusion was that population growth is not sufficient
to explain the data.} \cite{Nevzorov87,Yang75}: 
As the 
newly introduced RV's are indistinguishable from those that have
appeared at earlier times, the probability that a record occurs
among them is simply equal to 
\begin{equation}
\label{Pk}
P_k = \frac{\alpha_k}{\sum_{j=1}^k \alpha_j} \approx 1 - \frac{k}{(\ell - 1) (N-k)}.
\end{equation}
In the last step the expression (\ref{alphak}) has been inserted and
an expansion for $k, N \to \infty$ at fixed $k/N$ has been carried
out \cite{Jain04}. The probability $P_k$ starts out at unity and dwindles
to zero as $k$ approaches the value $k_\mathrm{max}$ of the Hamming distance
at which the majority of sequences reside; the process stops at
$t = k_\mathrm{max}$, when the globally fittest sequence $\sigma^{(f)}$ (which is located
with certainty at $k_\mathrm{max}$ for large $N$) is reached. 
In contrast to the logarithmic increase (\ref{log}) in the i.i.d. case,
here new records are found quite frequently, at least when $k \ll N$. 
This is because of the exponential
growth of the number of available sequences with increasing distance from
the seed, which compensates the scarcity of new records. 

Integrating
(\ref{Pk}) from $k=0$ to $k_\mathrm{max}$ one finds that the 
mean of the total number
of records $R$ that are encountered during the evolution is given by 
\begin{equation}
\label{MeanR}
\langle R \rangle = \left( 1 - \frac{\ln \ell}{\ell -1} \right) N.
\end{equation}
It can be shown that the occurrences of records are 
independent events in this model \cite{Nevzorov87,Jain04}, and hence the variance and higher
moments of $R$ can also be computed from the $P_k$. The variance is   
\begin{equation}
\label{varR}
\langle R^2 \rangle - \langle R \rangle^2 = \sum_{k = 1}^{N} 
P_k - P_k^2 \approx \frac{N}{\ell - 1} 
\left( \frac{\ell + 1}{\ell - 1} \ln \ell - 2 \right)
\end{equation}
for large $N$, which decreases with increasing $\ell$. Thus asymptotically
$R$ is a normal RV with fluctuations of order $\sqrt{N}$.         
In addition, analytic results for the 
the spacings between records are reported in \cite{Jain04}.

\section{Quasispecies evolution in the strong selection limit}

For a somewhat more realistic description of the population dynamics,
we turn to Eigen's quasispecies model \cite{Eigen71,Eigen89}, arguably
the simplest mathematical model that implements the basic
mechanisms of selection and mutation for a genetically heterogeneous
population on the level of the sequence space \cite{Baake99}.
The model was introduced to describe the population dynamics of asexually
reproducing entities like self-replicating macromolecules. It can
be applied whenever the population size is large, so that the number of individuals occupying a given
site in sequence space can be represented by a continuous variable. 
Because of the exponential proliferation of the number of sequences with
increasing $N$, real populations are very sparse in sequence space,
which severely limits the applicability of a continuum description. 
We nevertheless believe that it is important to first understand
the long time dynamics of sequence space evolution in the continuum setting,
before taking into account the effects of the discreteness of real populations.

In the quasispecies model, the population $Z(\sigma,t)$ of genotype $\sigma$
at time $t$ evolves in discrete time according to the linear recursion relation
\begin{equation}
\label{quasi}
Z(\sigma,t+1) = \sum_{\sigma'} p(\sigma' \to \sigma) W(\sigma') Z(\sigma',t),
\end{equation}
where $p(\sigma' \to \sigma)$ is the mutation probability that sequence $\sigma$ appears
as offspring of sequence $\sigma'$. Assuming that single point mutations occur
with probability $\mu$ per generation, the mutation probability takes the
form 
\begin{equation}
\label{mutation}
p(\sigma' \to \sigma) = \mu^{d(\sigma,\sigma')} (1- \mu)^{N-d(\sigma,\sigma')}.
\end{equation}

Consider a population that is initially localized at a seed sequence $\sigma_0$, 
i.e., the initial condition for (\ref{quasi}) is $Z(\sigma,0) = Z_0 \delta_{\sigma,\sigma_0}$.
Then after one time step we have 
\begin{equation}
\label{initial}
Z(\sigma,1) = Z_0 W(\sigma_0) (1-\mu)^N [\mu/(1-\mu)]^{d(\sigma,\sigma_0)} 
\sim \exp[-d(\sigma,\sigma_0)/\lambda].
\end{equation}
The population density is now nonzero everywhere, with a magnitude decaying exponentially
with increasing distance from the seed sequence, where the decay length is
$\lambda = 1/\ln[1/\mu-1]$. At this point individuals with genotypes far away from the seed start
to compete with the majority of the population still located at $\sigma_0$. To quantify
this competition, we follow the location of the \emph{current leader} $\sigma^\ast(t)$,
which is defined as the sequence at which $Z(\sigma,t)$ is maximal. The path of 
$\sigma^\ast(t)$ describes an \emph{evolutionary trajectory} in sequence space
\cite{Krug02,Krug03}. Along such a trajectory the fitness $F(\sigma^\ast)$
increases in a stepwise fashion, similar to the \textit{fitness trajectories} observed
in experimental studies of microbial populations \cite{Lenski94,Burch99}. 

The dynamics of evolutionary trajectories is simple in a \emph{strong
selection limit} modeled after the zero temperature limit of the statistical
physics of disordered systems \cite{Krug02,Krug03}. Writing 
$\mu = e^{-\beta \gamma}$ and taking the inverse selective 
temperature $\beta \to \infty$, one obtains 
a recursion relation for the logarithmic population variable $E(\sigma,t)$ defined by 
$Z(\sigma,t) =  e^{\beta E(\sigma,t)}$. As was shown in \cite{Krug03}, the behavior
remains essentially unchanged if the mutational part of the 
dynamics is turned off after the first time step. This implies that for $t \geq 2$ the population
at each site $\sigma$ grows independently, at its own logarithmic rate $F(\sigma)$,
according to 
\begin{equation}
\label{linear}
E(\sigma,t) = E(\sigma,1) + F(\sigma)(t-1) = F(\sigma_0) - \gamma d(\sigma,\sigma_0) + 
F(\sigma)(t-1).   
\end{equation}
Here the initial condition (\ref{initial}) has been inserted. 
Equation (\ref{linear}) is a particularly transparent representation of the 
evolutionary race. 
Each genotype advances at its own speed $F(\sigma)$, from an initial
position determined by its distance from the seed sequence $\sigma_0$.
In the course of time, the leadership in the population changes from sequences
with relatively low fitness located close to $\sigma_0$ to more distant sequences
of larger fitness, until eventually the globally fittest sequence is reached and the
race comes to an end. At any given time the current leader $\sigma^\ast(t)$ 
satisfies $E(\sigma^\ast(t),t) = \max_{\sigma} \{E(\sigma,t)\}$; that is, 
$E(\sigma^\ast(t),t)$ is the upper envelope of the family of straight
lines defined by (\ref{linear}), and \emph{leadership changes} correspond
to the corners of the envelope. 
The leadership changes are precisely the jumps in the punctuated
evolutionary trajectory, and their statistics will be discussed
in the next section. 

\section{Bypassing}
\label{Bypassing}

Several properties of evolutionary trajectories follow immediately from the
representation (\ref{linear}). First, since all sequences within a \textit{shell}
of constant distance $d(\sigma,\sigma_0)$ start with the same population at $t=1$,
only the sequence with the largest fitness within each shell has a chance
of ever attaining the leadership. Second, in order to become the new leader,
the fitness of a sequence has to exceed that of the current leader, i.e. the 
sequence has to be a \emph{record} in the sense of Sect.\ref{RecordsSeq}. 
Thus, among the $\ell^N$ available genotypes only a small fraction 
given by the mean number of records (\ref{MeanR}) is eligible to become
part of the evolutionary trajectory.

However, not every record will become a leader. To see this, suppose 
the current leader is at $\sigma$, and let $\sigma'$ be a subsequent record
with $F(\sigma') > F(\sigma)$ and $d(\sigma_0,\sigma') > d(\sigma_0,\sigma)$.
Then $E(\sigma,t)$ and $E(\sigma',t)$ will cross at time
\begin{equation}
\label{Tsigma}
T(\sigma,\sigma') = \frac{\gamma[d(\sigma_0,\sigma') - d(\sigma_0,\sigma)]}{F(\sigma')-
F(\sigma)}.
\end{equation}
The leadership will be taken over by the sequence $\sigma'$ that 
minimizes the crossing time (\ref{Tsigma}), which
does not need to be the next record in line. We say that a sequence $\sigma_1$ is 
\emph{bypassed} by a sequence $\sigma_2$ with $d(\sigma_0,\sigma)< d(\sigma_0,\sigma_1) < d(\sigma_0,\sigma_2)$,
if $T(\sigma,\sigma_2) < T(\sigma,\sigma_1)$. Because of bypassing, the number
of records (\ref{MeanR}) is only an upper bound on the number of leadership changes.

In contrast to the properties of the records discussed in Sects.\ref{Records} and
\ref{RecordsSeq}, which are independent of the underlying fitness distribution, 
the prevalence of bypassing depends on $p(F)$ \cite{Jain04}. We can get some
insight into the behavior by estimating the typical time $T^\ast = T(\sigma^{(f-1)},\sigma^{(f)})$
at which the penultimate leader $\sigma^{(f-1)}$ is overtaken by the sequence
$\sigma^{(f)}$ with globally maximal fitness \cite{Krug03}. As both $\sigma^{(f)}$
and $\sigma^{(f-1)}$ are expected to reside within a belt of thickness $\sqrt{N}$
around $k_{\mathrm{max}}$, we have $d(\sigma_0,\sigma^{(f)}) - d(\sigma_0,\sigma^{(f-1)}) \sim 
\sqrt{N}$. The fitness difference $F(\sigma^{(f)}) - F(\sigma^{f-1})$ should be 
of the order of the fitness gap of the landscape. 
For example, for 
a fitness distribution with a power law tail $p(F) \sim F^{-(\mu+1)}$,
we have according to (\ref{Fmax}) that $F_{\mathrm{max}} \sim S^{1/\mu} = \ell^{N/\mu}$, and 
the fitness gap is of the same order. This implies that the crossing time $T^\ast \sim \sqrt{N}/\ell^{N/\mu}$ 
decreases\footnote{Due to the rare occurrence of landscapes with a very small
fitness gap, the mean crossing time is nevertheless infinite: The distribution of
$T^\ast$ has a universal $1/(T^\ast)^2$ tail with a prefactor that vanishes
for $N \to \infty$ for power law fitness distributions \cite{Krug03}.} 
with increasing $N$; for large $N$ 
\emph{all} intermediate records are bypassed, and the globally fittest sequence 
immediately takes over the leadership. 

\begin{figure}
\centerline{\includegraphics[width=0.5\textwidth,angle=-90]{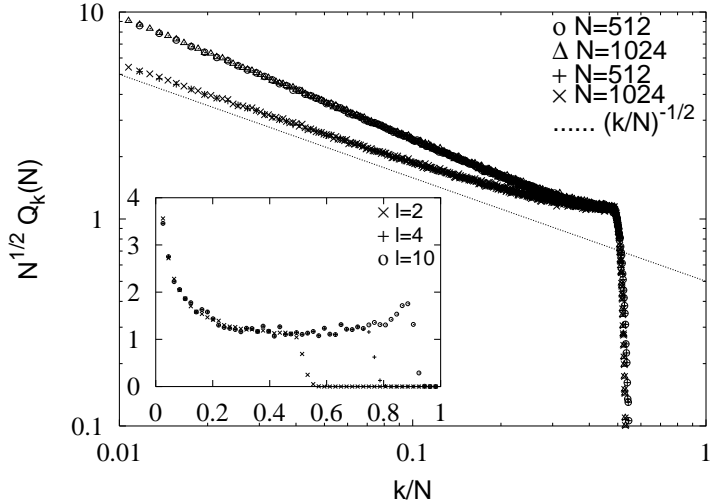}}
\caption{Simulation data for the probability $Q_k$ for an evolutionary
jump to occur at distance $k$ from the seed site. 
Main figure shows data for Gaussian ($\mathrm{O}, \Delta$) and exponential
($+, \times$) fitness distributions, two different sequence
lengths $N$, and alphabet size $\ell = 2$, on a double logarithmic scale.
Inset shows data for exponential fitness distribution, $N=512$, and three 
different values of $\ell$ on a linear scale. 
The data were averaged over $10^5$ (main figure)
and $10^4$ (inset) disorder configurations, respectively.}
\label{Shellfig}
\end{figure}

Nontrivial behavior is found when the fitness gap decreases with $N$, or increases more slowly than
$\sqrt{N}$.  
In Fig.\ref{Shellfig} we show numerical data for Gaussian and  
exponential fitness distributions, 
for which the gap is of order unity independent of $N$.
Very large sequence lengths can be treated by 
using the \textit{shell fitness} $F_k$, which is the largest of $\alpha_k$ 
i.i.d. RV's \cite{Krug03}; in this way the number of RV's that are needed for
each realization reduces from $\ell^N$ to $N$. The generation of the $F_k$ is feasible despite
the astronomically large values of $\alpha_k$ because the maximum of $\alpha_k$ exponential 
or Gaussian RV's
is only of order $\ln \alpha_k$ or $\sqrt{\ln \alpha_k}$, respectively
[compare to (\ref{Fmax})]. 
The key result illustrated in Fig.\ref{Shellfig} is the scaling form 
\begin{equation}
\label{Pktilde}
Q_k \approx N^{-1/2} f(k/N) 
\end{equation}
for the probability $Q_k$ for an evolutionary jump to occur at distance $k$ from the seed.
The total number of jumps is of order $\sqrt{N}$, and hence
most of the $O(N)$ records are bypassed.
The scaling function $f(x)$ is cut off at $k_{\mathrm{max}}/N = 1 - 1/\ell$, but its
shape appears to be independent of the alphabet size $\ell$ (inset of Fig.\ref{Shellfig}).
For both Gaussian and exponential fitness distributions, the behavior of the scaling
function at small arguments is close to $f(x) \sim x^{-1/2}$. This behavior would imply 
that $Q_k \sim 1/\sqrt{k}$ independent of $N$ for 
$k \ll N$, and that the number of jumps grows as $\sqrt{k}$ with increasing distance
from the seed site. 
We expect these results to be generally valid
for fitness distributions in the Gumbel universality class of extreme value
theory \cite{Sornette00}. 
The case of bounded fitness distributions should also be interesting,
but has not been treated so far because of the difficulty in creating the
shell fitnesses for large $N$.
  
An analytic understanding of (\ref{Pktilde}) is lacking at present, and must be left
to future work. In fact, as is explained in detail in \cite{Jain04}, the statistics
of bypassing is difficult to handle analytically even for the simple case when the
geometry of sequence space is ignored and the shell fitnesses are replaced by i.i.d. RV's. 
It is remarkable that the innocuous generalization of the basic record model, defined by
the family (\ref{linear}) of lines with random slopes, leads to a rather involved and 
rich probabilistic problem.

\section*{Acknowledgements}

We acknowledge useful discussions with Andreas Engel and Luca Peliti.
This work has been supported by DFG within SFB/TR 12 \textit{Symmetries and Universality
in Mesoscopic Systems}.


\begin{thebibliography}{00}
\bibitem{Darwin} C. Darwin, The Origin of Species by Means of Natural Selection, 
John Murray, London, 1859.
\bibitem{Kauffman87} S.A. Kauffman, S. Levin, J. theor. Biol. 
\textbf{128} (1987) 11.
\bibitem{Sibani98} P. Sibani, M. Brandt, P. Alstr\o m, Int. J. Mod. 
Phys. {\bf 12} (1998) 361.
\bibitem{Glick78} N. Glick, Amer. Math. Monthly {\bf 85} (1978) 2. 
\bibitem{Nevzorov87} V. B. Nevzorov, Theory Probab. Appl. {\bf 32} (1987) 201.
\bibitem{Feller71} W. Feller, Introduction to Probability Theory and
Its Applications, Vol.2, Wiley, New York, 1971.  
\bibitem{Newman99} M.E.J. Newman, P. Sibani, Proc. Roy. Soc. London B {\bf 266} (1999) 1593.
\bibitem{Sibani03} P. Sibani, J. Dall, Europhys. Lett. \textbf{64} (2003) 8.
\bibitem{Gould93} S.J. Gould, N. Eldredge, Nature {\bf 366} (1993) 223.
\bibitem{Eldredge89} N. Eldredge, Macroevolutionary Dynamics, McGraw-Hill, New York, 1989.
\bibitem{Krug02} J. Krug, in:  M. L\"assig, A. Valleriani (Eds.), 
Biological Evolution and Statistical Physics, Springer, Berlin, 2002, p. 205.
\bibitem{Krug03} J. Krug, C. Karl, Physica A {\bf 318} (2003) 137.
\bibitem{Jain04} K. Jain, J. Krug (to be published).
\bibitem{Baake99} E. Baake, W. Gabriel,  
in: D. Stauffer (Ed.), Annual Reviews of Computational Physics VII,
World Scientific, Singapore, 2000,
pp. 203-264. 
\bibitem{Peliti95} L. Peliti, in: T. Riste, D. Sherrington (Eds.),
Physics of Biomaterials: Fluctuations, Self-Assembly and Evolution,
Kluwer, Dordrecht, 1996, p. 267.
\bibitem{Peliti97} L. Peliti, {\tt{cond-mat/9712027}}.
\bibitem{Franz93} S. Franz, M. Sellitto, L. Peliti, J. Phys. A \textbf{26} (1993) L1195. 
\bibitem{Flyvbjerg92} H. Flyvbjerg, B. Lautrup, Phys. Rev. A
{\bf 46} (1992) 6714.
\bibitem{David70} H.A. David, Order Statistics, Wiley, New York, 1970.
\bibitem{Sornette00} D. Sornette, Critical Phenomena in Natural Sciences, 
Springer, Berlin, 2000.
\bibitem{Yang75} M.C.K. Yang, J. Appl. Prob. {\bf 12} (1975) 148.
\bibitem{Eigen71} M. Eigen, Naturwissenschaften {\bf 58} (1971) 465. 
\bibitem{Eigen89}
M. Eigen, J. McCaskill, P. Schuster, Adv. Chem. Phys. {\bf 75} (1989) 149.
\bibitem{Lenski94} R.E. Lenski, M. Travisano, Proc. Natl. Acad. Sci. USA 
\textbf{91} (1994) 6808.
\bibitem{Burch99} C.L. Burch, L. Chao, Genetics \textbf{151} (1999) 921.

\end{thebibliography}
\end{document}